\documentclass[prl, twocolumn, floatfix]{revtex4}

\usepackage{graphicx}
\usepackage{wasysym}

\begin{document}

\title{Spin-Polarized Transient Electron Trapping in Phosphorus-doped Silicon}

\author{Yuan Lu}
\altaffiliation{Present address: CNRS Institut Jean Lamour, Nancy, FR}
\author{Jing Li}
\author{Ian Appelbaum}
\altaffiliation{appelbaum@physics.umd.edu}
\affiliation{Department of Physics and Center for Nanophysics and Advanced Materials, University of Maryland, College Park MD 20742 USA}

\begin{abstract}
Experimental evidence of electron spin precession during travel through the phosphorus-doped Si channel of an all-electrical device simultaneously indicates two distinct processes: \emph{(i)} short timescales ($\approx$50ps) due to purely conduction-band transport from injector to detector and \emph{(ii)} long timescales ($\approx$1ns) originating from delays associated with capture/re-emission in shallow impurity traps. The origin of this phenomenon, examined via temperature, voltage, and electron density dependence measurements, is established by means of comparison to a numerical model and is shown to reveal the participation of metastable excited states in the phosphorus impurity spectrum. This work therefore demonstrates the potential to make the study of macroscopic spin transport relevant to the quantum regime of individual spin interactions with impurities as envisioned for quantum information applications.  
\end{abstract}

\maketitle

Incorporating impurities into the otherwise pure silicon (Si) lattice has numerous consequences for charge transport. Most importantly, doping using atomic species with more than Si's four valence electrons not only provides mobile charge to the conduction band, but also leaves behind a positively-charged ion that modifies the electrostatic energy landscape when not screened. But beyond impacting the flow of electron charge, electron (or n-type) doping also impacts spin-polarized transport as well.\cite{JANSENJONKER, JONKERLATERAL, SHIRAISHIHANLE}  

For example, several years ago we showed how band bending gives rise to non-ohmic spin transport in $\approx$3 $\mu$m-thick n-type lightly phosphorus-doped Si using all-electrical ballistic hot electron injection and detection techniques.\cite{DOPED} In the present work, we show that under certain circumstances, long-timescale processes not seen in otherwise-equivalent undoped devices\cite{APPELBAUMNATURE, BIQINPRL} can be observed in measurements of devices with lowly-doped transport channel regions. These results are attributed to the interaction of conduction electrons with shallow impurity-related traps and suggest that the study of spin-polarized transport in semiconductors can potentially be used to elucidate physics previously accessible only to time-domain techniques or to explore the local interaction of spin information with isolated impurity potentials.

\begin{figure}
\includegraphics[width=7cm, height=3.75cm]{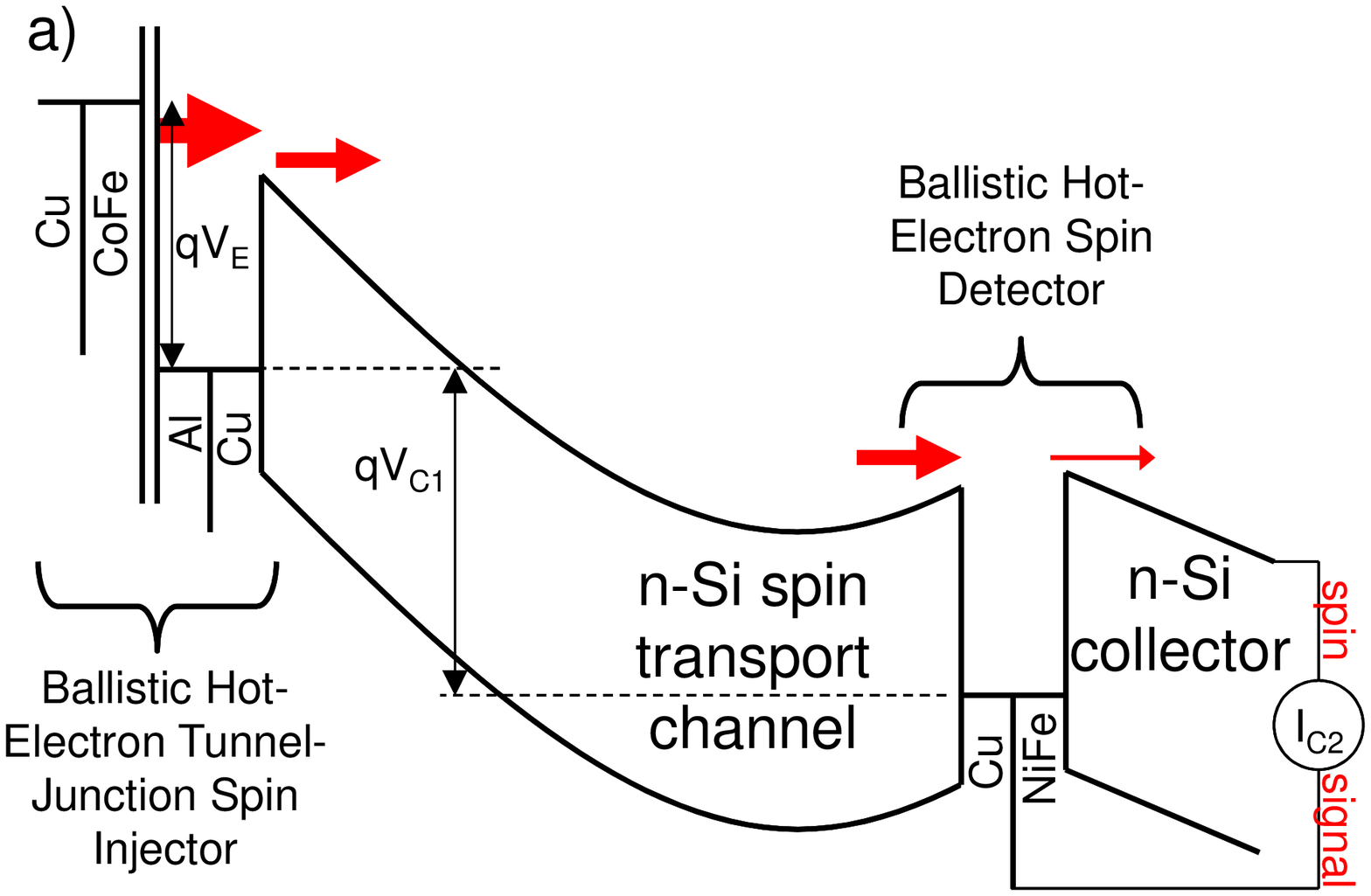}
\includegraphics[width=6cm, height=2.5cm]{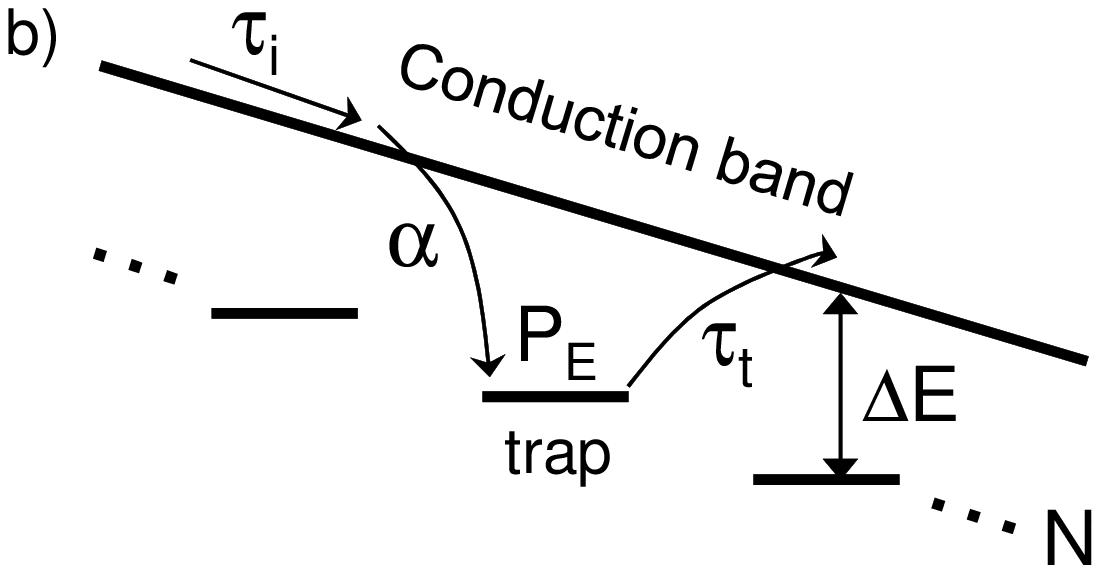}
\caption{\label{BANDFIG}
(a) Schematic band diagram of vertical n-type doped Si spin transport channel devices used in this work, showing band-bending and the resultant potential minimum for temperatures high enough to ionize dopants and low applied voltage $V_{C1}$. (b) Illustration of model implied by Eqn. \ref{DISTREQN} for capture of electrons (incident with period $\tau_i$) by $N$ shallow traps $\Delta$E below the conduction band with probability $\alpha$. Each trap is initially empty with probability $P_E$ and releases the electron with an exponential probability distribution having timescale $\tau_t$.
}
\end{figure}

The devices we use here are nominally identical to those in Ref. \onlinecite{DOPED}, but with slightly lower impurity density. As illustrated in Fig. \ref{BANDFIG} (a), spin-polarized hot electrons are injected from a ferromagnetic (FM) tunnel-junction cathode (CoFe, biased at emitter voltage $V_E$) and travel ballistically through a nonmagnetic thin film and over an energy-filtering Schottky barrier to couple with conduction band states in the 3-micron-thick n-Si transport layer. Spin detection after vertical transport through this layer (biased at collector voltage $V_{C1}$) is accomplished by analyzing the ballistic component of the hot-electron current generated by ejection from the n-Si conduction band over a Schottky barrier and into a second FM thin film (NiFe). Because the mean-free-path of hot electrons is determined by the relative orientation between their spin and the magnetization of the FM film, the magnitude of this current ($I_{C2}$) then carries information about spin transport in the channel. 

Previous measurements on these devices have indicated that the confluence of Schottky depletion regions on both injector and detector sides of the transport channel result in a confining conduction-band profile (i.e. an electrostatic potential energy minimum exists) for applied voltages between injector and detector ($V_{C1}$) below $\approx$2 V at temperatures $\apprge$ 40 K, sufficient to ionize the dopants.\cite{YUAN} Assuming full depletion, this biasing behavior corresponds to a doping density of approximately 3$\times$10$^{14}$ cm$^{-3}$.

\begin{figure}
\includegraphics[width=6cm, height=8cm]{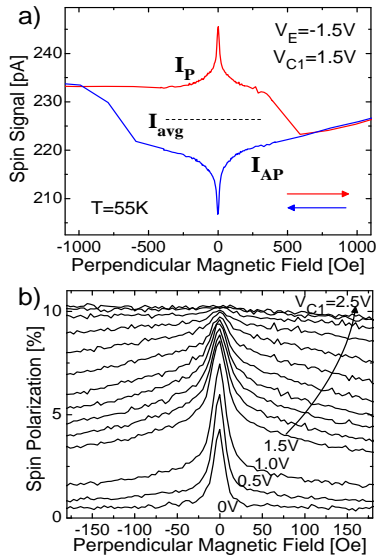}
\caption{\label{VOLTFIG}
Perpendicular-field measurements at 55K. In (a), a full field loop displays square hysteresis and clear low-field dephasing. In (b), voltage dependence of the low-field signal is shown, where the applied voltage $V_{c1}$ is changed from 0V, 0.5V, 1.0V, and 1.5V to 2.5V in steps of 0.1V. 
}
\end{figure}

In a magnetic field oriented perpendicular to the device plane, the magnetization of injector and detector magnetic thin films remains in-plane due to shape anisotropy. However, spins injected into the nonmagnetic Si are induced to precess around this magnetic field; in drift-dominated transport through the conduction band, the transit time is well-defined and this leads to coherent spin-signal oscillations as a function of magnetic field magnitude.\cite{BIQINPRL, DEPHASINGPRB, LARMORPRB} On the other hand, when a large transit-time uncertainty exists (due e.g. to random diffusion), electron spins arrive at the detector with a distribution of precession angles. This spin ``dephasing'' suppresses the oscillation amplitude. A Fourier-transform method applied to the quasi-statically measured magnetic-field spectroscopy can directly recover the empirical spin transit-time distribution with ns resolution\cite{LATERAL, LARMORPRB} and indicates an inverse relationship between the magnitude of magnetic-field features and transport timescales.    

Measurements on our phosphorus-doped Si devices with a perpendicular magnetic field (at T=55K in Fig. \ref{VOLTFIG} (a)) show an open loop corresponding to in-plane magnetization switching of the softer NiFe detector layer (induced by small field misalignments\cite{OBLIQUE}), superimposed on the signal due to spin precession during transport. It is evident that our spin transport signal includes the signatures of two \emph{distinct} processes: both (i) precession features on the kOe scale beyond the range shown due to the short ($\approx$50ps) timescale associated with conduction-band transport from injector to detector, and (ii) a low-field dephasing resulting in a narrow peak of width 50-100 Oe due to much longer timescales. This observation is further supported by noticing that the small-field peak in a parallel injector-detector magnetization configuration (left-to-right sweep, red) becomes inverted for the antiparallel configuration (right-to-left sweep, blue).

This small-field peak is sensitive to the applied voltage between injector and detector ($V_{C1}$). In Fig. \ref{VOLTFIG}(b), we show the low-field region of our measurements from parallel magnetization ($I_P$), where changes in absolute signal current as a function of applied voltage are eliminated by plotting spin polarization $(I_P-I_{avg})/I_{avg}$, where $I_{avg}$ is the average between parallel and anti-parallel signals, as shown by the horizontal dotted line in Fig. \ref{VOLTFIG}(a). Clearly, higher applied voltage corresponds to a widening peak that accounts for less of the total spin polarization signal. At voltages sufficient to entirely remove the confining potential in the conduction band  ($\approx V_{C1}=$2.5V), the low-field peak disappears entirely. 

\begin{figure}
\includegraphics[width=6cm, height=8cm]{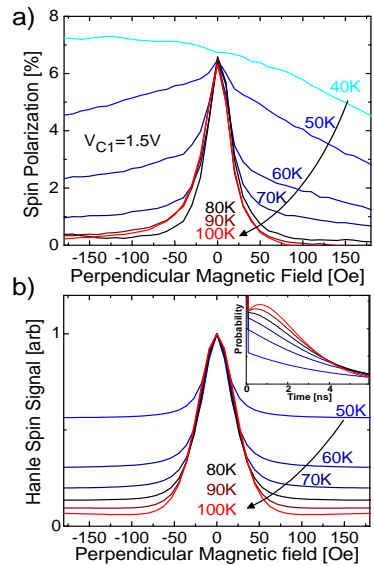}
\caption{\label{TEMPFIG}
Temperature dependence of low-field dephasing between 40K and 100K at applied voltage $V_{C1}=$1.5V. (b) Simulated spin-precession Hanle signals using the trapping-time model (Eq. \ref{DISTREQN}) with parameters $\alpha$=0.2, $\tau_0$=200ps, $\Delta E$=11.5meV, and effective 1-d current of 3nA sampling $N=$50 traps. Inset: Calculated trapping-time distributions.} 
\end{figure}

Temperature dependence of this phenomenon at a fixed $V_{C1}=$1.5V is shown in Fig. \ref{TEMPFIG}(a). Here, we see that increasing temperature reduces the relative contribution of the low-field peak to the overall signal and obscures the simultaneous presence of large-field scales. Observation of the dual-timescale phenomena it otherwise implies therefore requires a narrow range of both voltages and temperatures; this explains why it was overlooked during the study which led to Ref. \onlinecite{DOPED}.

This low-field feature is reminiscent of similar observations in optical\cite{PAGET} and transport\cite{CHANCROWELL} measurements of spin-polarized electron precession in GaAs. In that case, the phenomena was attributed to a non-linear dynamic feedback between electron spin and nuclear spin via Overhauser and Knight fields.\cite{OVERHAUSER} Ref. \onlinecite{CHANCROWELL}, for instance, demonstrates a remarkable degree of correspondence with steady-state finite-difference time-domain calculations of spin transport in the presence of the self-consistent nuclear field, and its modification by microwave excitation resonant with the nuclear spin splitting. 

Despite the similarity to the results observed here with Si, however, there are several reasons to disregard this hyperfine-mediated mechanism as a possible origin. Si is expected to have much weaker hyperfine interactions than GaAs due to the small relative abundance of nonzero nuclear spin isotopes ($\approx$5\% for spin-1/2 $^{29}$Si; spin-1/2 $^{31}$P dopants are even more dilute). In comparison, every nucleus in GaAs has nonzero spin. Furthermore, we do not observe the measurement time dependence associated with long nuclear spin lifetimes seen in GaAs spin transport devices.\cite{CHANCROWELL, SALISHYPERFINE} 

Our spin current signal is essentially a measurement of the average spin precession angle $\theta$, which is a product of both spin precession frequency $\omega$ and transit time $t$. In principle, the effects discussed above could be the result of modification of either of these; a linear process would require that there is a sub-ensemble of electrons either moving substantially slower or precessing substantially faster than the rest to account for both large- and small-field features. The latter cause is eliminated by the nonexistence of g-factors significantly greater than 2 in Si.

The origin of our observations is therefore a transit-time effect. Here, we establish that this is due to the presence of trapping into and subsequent re-emission of spin-polarized conduction electrons from shallow phosphorus impurity-related states. At typical injection currents of $\approx$10 $\mu$A, there are approximately $10^{14}$ electrons entering (and leaving) the transport channel each second. For a quasi-Lorentzian Hanle half-width at half-maximum of $\Delta B\approx$50 Oe, an average transit time of $\hbar/(g\mu_B \Delta B)\approx$1 ns implies a steady-state population of 10$^5$ occupied traps in the channel. This value is below the actual density of approximately $10^7$ available charged impurities under conditions above the ionization temperature in the effective volume $\approx$100$\mu$m$\times$100$\mu$m$\times$3$\mu$m at the phosphorus density used here (although this number must be reduced at low temperatures due to non-negligible occupation probability). 
  
\begin{figure}
\includegraphics[width=6cm, height=8cm]{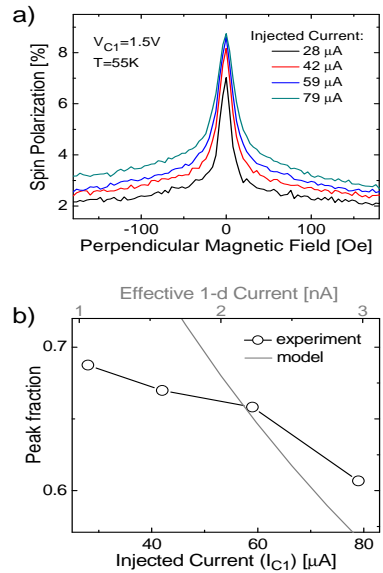}
\caption{\label{CURRFIG}
Low-field dephasing as a function of injected current and hence steady-state electron density. In (a), spin polarization is shown, whereas (b) demonstrates a reduction in relative contribution to the overall signal from trapped electrons when density increases. Our numerical model included for comparison uses an effective 1-dimensional current. }
\end{figure}

By varying the electron density via injected current ($I_{C1}$), we can see the signature of this fixed trap number. Although the available range of this parameter is limited by the reliable operating conditions of our tunnel junction spin injector, one can clearly see in Fig. \ref{CURRFIG}(a) that the overall polarization increases with injected current. The relative contribution of the low-field peak from this data is shown in Fig. \ref{CURRFIG}(b), from which it is apparent that a small but substantial decrease in the relative proportion of trapped electrons (i.e. relative peak height to background polarization) accompanies an increase in conduction electrons as well. This behavior indicates the approach to saturation of the impurities and a fixed contribution to low-field dephasing.  

Average transit times for electrons participating in trapping events of approximately 1ns (indicated by the width of the low-field peak) are relatively small at these temperatures for the expected ground-state phosphorus-impurity trap depth of 45meV.\cite{SZEBOOK} Consistency of the present trapping scheme with the observed timescales requires the participation of trapping-emission cycles involving shallow states that are presumably excited Rydberg states with suppressed relaxation to the relatively deep ground state. In this regard, it is supportive to note that recent time-domain experiments with a far-infrared pulsed free-electron laser has identified transition lifetimes of the phosphorus 2p$_0$ excited state ($\Delta E=$11.5meV below the conduction band) exceeding 200ps in this temperature range.\cite{AEPPLI} Furthermore, recent calculations suggest an intrinsic 2p-1s transition timescale of over 1ns,\cite{2P0CALC} which has even enabled population inversion and lasing in externally-pumped systems.\cite{THZLASER} 

We now describe a simple numerical model to calculate the transit-time probability function which incorporates the details of trapping into (and emission out of) the impurity state. To capture the observed behavior, one must consider the convolution of all $0\leq k \leq N$ possible trapping events in $N$ available traps with trapping probability $\alpha$: 

\begin{equation}
(1-\alpha P_E)^N\delta(t)+\sum_{k=1}^N \left(^N_k \right)(1-\alpha P_E)^{N-k}(\alpha P_E)^k G(t,k,\tau_t)
\label{DISTREQN}
\end{equation}

\noindent where $G(t,k,\tau_t)$ is the gamma probability distribution (convolution of $k$ exponentials of trapping timescale $\tau_t$). The probability of an empty trap ($P_E$) is given by the relative ratio of incident timescale ($\tau_i$ determined by the injection current), to trapping timescale ($\tau_t=\tau_0e^{\Delta E/k_B T}$) such that $P_E=1-\frac{\tau_t}{\tau_i/\alpha+\tau_t}$. The physical significance of the relevant parameters of this model are illustrated in Fig. \ref{BANDFIG} (b).

Since this a one-dimensional model, an equivalent ``effective'' current (to determine $\tau_i$) and trap number ($N$) must be used. At the phosphorus density used here, the inter-dopant distance is approximately 150nm. Therefore, with a total electron injection area of 10$^4 \mu$m$^2$, $\approx 2.25\times 10^{-6}$ of the injected current will interact with $N\approx 20$ dopants through the 3-$\mu$m transit length. A real injected current $I_{C1}=79\mu$A at $V_E=-1.5$V is therefore equivalent to an effective 1-d current of approximately 200pA. Allowing for some degree of variation in these parameters given the simplicity of the analogy, we use an effective current of 3nA and $N=50$ in our subsequent comparison of the model fit to the experimental data.

Ignoring the contribution from the much smaller timescale features endowed by drift/diffusion/relaxation, the observed spin signal is then the real part of the Fourier transform of the distribution in Eqn. \ref{DISTREQN}.\cite{LATERAL, LARMORPRB} Correspondence of this model's numerical predictions to the general features of experimental data are evident in Fig. \ref{TEMPFIG}(b), where the changes in trapping contribution and weak linewidth dependence on temperature observed in experimental results in Fig. \ref{TEMPFIG}(a) are clearly reproduced. Furthermore, we also show that the numerical model captures the dependence of the relative contribution from trapped electrons on injected current as seen in Fig. \ref{CURRFIG} (b) using identical model parameters. Although we have not included it in the model, the broadening and peak reduction seen in voltage dependence (Fig. \ref{VOLTFIG}) can be attributed to field emission and a subsequent reduction in trapping timescale $\tau_t$ and capture probability $\alpha$.

The explicit coupling of spin-polarized conduction electrons with phosphorus impurities as demonstrated here opens many research possibilities. It may be useful as a probe of impurity levels and transition rates not explicitly requiring time-domain methods as previously assumed were necessary.\cite{AEPPLI} In addition, we speculate that it may become possible to explore the role of contact hyperfine interactions and dynamic nuclear polarization of the phosphorus nuclear spin by itinerant non-equilibrium spin-polarized conduction electrons with NMR techniques for potential application to quantum computing schemes.\cite{ITOHNMR} 

We acknowledge helpful comments by S. Bohacek. This work was supported by the Office of Naval Research and the National Science Foundation. We acknowledge the support of the Maryland NanoCenter and its FabLab.


\begin{thebibliography}{20}
\expandafter\ifx\csname natexlab\endcsname\relax\def\natexlab#1{#1}\fi
\expandafter\ifx\csname bibnamefont\endcsname\relax
  \def\bibnamefont#1{#1}\fi
\expandafter\ifx\csname bibfnamefont\endcsname\relax
  \def\bibfnamefont#1{#1}\fi
\expandafter\ifx\csname citenamefont\endcsname\relax
  \def\citenamefont#1{#1}\fi
\expandafter\ifx\csname url\endcsname\relax
  \def\url#1{\texttt{#1}}\fi
\expandafter\ifx\csname urlprefix\endcsname\relax\def\urlprefix{URL }\fi
\providecommand{\bibinfo}[2]{#2}
\providecommand{\eprint}[2][]{\url{#2}}

\bibitem[{\citenamefont{Jansen et~al.}(2010)\citenamefont{Jansen, Min, Dash,
  Sharma, Kioseoglou, Hanbicki, van~'t Erve, Thompson, and
  Jonker}}]{JANSENJONKER}
\bibinfo{author}{\bibfnamefont{R.}~\bibnamefont{Jansen}},
  \bibinfo{author}{\bibfnamefont{B.~C.} \bibnamefont{Min}},
  \bibinfo{author}{\bibfnamefont{S.~P.} \bibnamefont{Dash}},
  \bibinfo{author}{\bibfnamefont{S.}~\bibnamefont{Sharma}},
  \bibinfo{author}{\bibfnamefont{G.}~\bibnamefont{Kioseoglou}},
  \bibinfo{author}{\bibfnamefont{A.~T.} \bibnamefont{Hanbicki}},
  \bibinfo{author}{\bibfnamefont{O.~M.~J.} \bibnamefont{van~'t Erve}},
  \bibinfo{author}{\bibfnamefont{P.~E.} \bibnamefont{Thompson}},
  \bibnamefont{and} \bibinfo{author}{\bibfnamefont{B.~T.}
  \bibnamefont{Jonker}}, \bibinfo{journal}{Phys. Rev. B}
  \textbf{\bibinfo{volume}{82}}, \bibinfo{pages}{241305}
  (\bibinfo{year}{2010}).

\bibitem[{\citenamefont{van~'t Erve et~al.}(2007)\citenamefont{van~'t Erve,
  Hanbicki, Holub, Li, Awo-Affouda, Thompson, and Jonker}}]{JONKERLATERAL}
\bibinfo{author}{\bibfnamefont{O.}~\bibnamefont{van~'t Erve}},
  \bibinfo{author}{\bibfnamefont{A.}~\bibnamefont{Hanbicki}},
  \bibinfo{author}{\bibfnamefont{M.}~\bibnamefont{Holub}},
  \bibinfo{author}{\bibfnamefont{C.}~\bibnamefont{Li}},
  \bibinfo{author}{\bibfnamefont{C.}~\bibnamefont{Awo-Affouda}},
  \bibinfo{author}{\bibfnamefont{P.}~\bibnamefont{Thompson}}, \bibnamefont{and}
  \bibinfo{author}{\bibfnamefont{B.}~\bibnamefont{Jonker}},
  \bibinfo{journal}{Appl. Phys. Lett.} \textbf{\bibinfo{volume}{91}},
  \bibinfo{pages}{212109} (\bibinfo{year}{2007}).

\bibitem[{\citenamefont{Sasaki et~al.}(2010)\citenamefont{Sasaki, Oikawa,
  Suzuki, Shiraishi, Suzuki, and Noguchi}}]{SHIRAISHIHANLE}
\bibinfo{author}{\bibfnamefont{T.}~\bibnamefont{Sasaki}},
  \bibinfo{author}{\bibfnamefont{T.}~\bibnamefont{Oikawa}},
  \bibinfo{author}{\bibfnamefont{T.}~\bibnamefont{Suzuki}},
  \bibinfo{author}{\bibfnamefont{M.}~\bibnamefont{Shiraishi}},
  \bibinfo{author}{\bibfnamefont{Y.}~\bibnamefont{Suzuki}}, \bibnamefont{and}
  \bibinfo{author}{\bibfnamefont{K.}~\bibnamefont{Noguchi}},
  \bibinfo{journal}{Appl. Phys. Lett.} \textbf{\bibinfo{volume}{96}},
  \bibinfo{pages}{122101} (\bibinfo{year}{2010}).

\bibitem[{\citenamefont{Jang et~al.}(2008)\citenamefont{Jang, Xu, Li, Huang,
  and Appelbaum}}]{DOPED}
\bibinfo{author}{\bibfnamefont{H.-J.} \bibnamefont{Jang}},
  \bibinfo{author}{\bibfnamefont{J.}~\bibnamefont{Xu}},
  \bibinfo{author}{\bibfnamefont{J.}~\bibnamefont{Li}},
  \bibinfo{author}{\bibfnamefont{B.}~\bibnamefont{Huang}}, \bibnamefont{and}
  \bibinfo{author}{\bibfnamefont{I.}~\bibnamefont{Appelbaum}},
  \bibinfo{journal}{Phys. Rev. B} \textbf{\bibinfo{volume}{78}},
  \bibinfo{pages}{165329} (\bibinfo{year}{2008}).

\bibitem[{\citenamefont{Appelbaum et~al.}(2007)\citenamefont{Appelbaum, Huang,
  and Monsma}}]{APPELBAUMNATURE}
\bibinfo{author}{\bibfnamefont{I.}~\bibnamefont{Appelbaum}},
  \bibinfo{author}{\bibfnamefont{B.}~\bibnamefont{Huang}}, \bibnamefont{and}
  \bibinfo{author}{\bibfnamefont{D.~J.} \bibnamefont{Monsma}},
  \bibinfo{journal}{Nature} \textbf{\bibinfo{volume}{447}},
  \bibinfo{pages}{295} (\bibinfo{year}{2007}).

\bibitem[{\citenamefont{Huang et~al.}(2007)\citenamefont{Huang, Monsma, and
  Appelbaum}}]{BIQINPRL}
\bibinfo{author}{\bibfnamefont{B.}~\bibnamefont{Huang}},
  \bibinfo{author}{\bibfnamefont{D.~J.} \bibnamefont{Monsma}},
  \bibnamefont{and}
  \bibinfo{author}{\bibfnamefont{I.}~\bibnamefont{Appelbaum}},
  \bibinfo{journal}{Phys. Rev. Lett.} \textbf{\bibinfo{volume}{99}},
  \bibinfo{pages}{177209} (\bibinfo{year}{2007}).

\bibitem[{\citenamefont{Lu and Appelbaum}(2010)}]{YUAN}
\bibinfo{author}{\bibfnamefont{Y.}~\bibnamefont{Lu}} \bibnamefont{and}
  \bibinfo{author}{\bibfnamefont{I.}~\bibnamefont{Appelbaum}},
  \bibinfo{journal}{Appl. Phys. Lett.} \textbf{\bibinfo{volume}{97}},
  \bibinfo{pages}{162501} (\bibinfo{year}{2010}).

\bibitem[{\citenamefont{Huang and Appelbaum}(2010)}]{LARMORPRB}
\bibinfo{author}{\bibfnamefont{B.}~\bibnamefont{Huang}} \bibnamefont{and}
  \bibinfo{author}{\bibfnamefont{I.}~\bibnamefont{Appelbaum}},
  \bibinfo{journal}{Phys. Rev. B} \textbf{\bibinfo{volume}{82}},
  \bibinfo{pages}{241202} (\bibinfo{year}{2010}).

\bibitem[{\citenamefont{Huang and Appelbaum}(2008)}]{DEPHASINGPRB}
\bibinfo{author}{\bibfnamefont{B.}~\bibnamefont{Huang}} \bibnamefont{and}
  \bibinfo{author}{\bibfnamefont{I.}~\bibnamefont{Appelbaum}},
  \bibinfo{journal}{Phys. Rev. B} \textbf{\bibinfo{volume}{77}},
  \bibinfo{pages}{165331} (\bibinfo{year}{2008}).

\bibitem[{\citenamefont{Jang and Appelbaum}(2009)}]{LATERAL}
\bibinfo{author}{\bibfnamefont{H.-J.} \bibnamefont{Jang}} \bibnamefont{and}
  \bibinfo{author}{\bibfnamefont{I.}~\bibnamefont{Appelbaum}},
  \bibinfo{journal}{Phys. Rev. Lett.} \textbf{\bibinfo{volume}{103}},
  \bibinfo{pages}{117202} (\bibinfo{year}{2009}).

\bibitem[{\citenamefont{Li et~al.}(2008)\citenamefont{Li, Huang, and
  Appelbaum}}]{OBLIQUE}
\bibinfo{author}{\bibfnamefont{J.}~\bibnamefont{Li}},
  \bibinfo{author}{\bibfnamefont{B.}~\bibnamefont{Huang}}, \bibnamefont{and}
  \bibinfo{author}{\bibfnamefont{I.}~\bibnamefont{Appelbaum}},
  \bibinfo{journal}{Appl. Phys. Lett.} \textbf{\bibinfo{volume}{92}},
  \bibinfo{pages}{142507} (\bibinfo{year}{2008}).

\bibitem[{\citenamefont{Paget et~al.}(1977)\citenamefont{Paget, Lampel,
  Sapoval, and Safarov}}]{PAGET}
\bibinfo{author}{\bibfnamefont{D.}~\bibnamefont{Paget}},
  \bibinfo{author}{\bibfnamefont{G.}~\bibnamefont{Lampel}},
  \bibinfo{author}{\bibfnamefont{B.}~\bibnamefont{Sapoval}}, \bibnamefont{and}
  \bibinfo{author}{\bibfnamefont{V.~I.} \bibnamefont{Safarov}},
  \bibinfo{journal}{Phys. Rev. B} \textbf{\bibinfo{volume}{15}},
  \bibinfo{pages}{5780} (\bibinfo{year}{1977}).

\bibitem[{\citenamefont{Chan et~al.}(2009)\citenamefont{Chan, Hu, Zhang, Kondo,
  Palmstr\o{}m, and Crowell}}]{CHANCROWELL}
\bibinfo{author}{\bibfnamefont{M.~K.} \bibnamefont{Chan}},
  \bibinfo{author}{\bibfnamefont{Q.~O.} \bibnamefont{Hu}},
  \bibinfo{author}{\bibfnamefont{J.}~\bibnamefont{Zhang}},
  \bibinfo{author}{\bibfnamefont{T.}~\bibnamefont{Kondo}},
  \bibinfo{author}{\bibfnamefont{C.~J.} \bibnamefont{Palmstr\o{}m}},
  \bibnamefont{and} \bibinfo{author}{\bibfnamefont{P.~A.}
  \bibnamefont{Crowell}}, \bibinfo{journal}{Phys. Rev. B}
  \textbf{\bibinfo{volume}{80}}, \bibinfo{pages}{161206}
  (\bibinfo{year}{2009}).

\bibitem[{\citenamefont{Overhauser}(1953)}]{OVERHAUSER}
\bibinfo{author}{\bibfnamefont{A.~W.} \bibnamefont{Overhauser}},
  \bibinfo{journal}{Phys. Rev.} \textbf{\bibinfo{volume}{92}},
  \bibinfo{pages}{411} (\bibinfo{year}{1953}).

\bibitem[{\citenamefont{Salis et~al.}(2009)\citenamefont{Salis, Fuhrer, and
  Alvarado}}]{SALISHYPERFINE}
\bibinfo{author}{\bibfnamefont{G.}~\bibnamefont{Salis}},
  \bibinfo{author}{\bibfnamefont{A.}~\bibnamefont{Fuhrer}}, \bibnamefont{and}
  \bibinfo{author}{\bibfnamefont{S.~F.} \bibnamefont{Alvarado}},
  \bibinfo{journal}{Phys. Rev. B} \textbf{\bibinfo{volume}{80}},
  \bibinfo{pages}{115332} (\bibinfo{year}{2009}).

\bibitem[{\citenamefont{Sze}(1981)}]{SZEBOOK}
\bibinfo{author}{\bibfnamefont{S.}~\bibnamefont{Sze}},
  \emph{\bibinfo{title}{Physics of Semiconductor Devices, 2nd edition}}
  (\bibinfo{publisher}{Wiley-Interscience}, \bibinfo{address}{New York},
  \bibinfo{year}{1981}).

\bibitem[{\citenamefont{Vinh et~al.}(2008)\citenamefont{Vinh, Greenland,
  Litvinenko, Redlich, van~der Meer, Lynch, Warner, Stoneham, Aeppli, Paul
  et~al.}}]{AEPPLI}
\bibinfo{author}{\bibfnamefont{N.}~\bibnamefont{Vinh}},
  \bibinfo{author}{\bibfnamefont{P.}~\bibnamefont{Greenland}},
  \bibinfo{author}{\bibfnamefont{K.}~\bibnamefont{Litvinenko}},
  \bibinfo{author}{\bibfnamefont{B.}~\bibnamefont{Redlich}},
  \bibinfo{author}{\bibfnamefont{A.}~\bibnamefont{van~der Meer}},
  \bibinfo{author}{\bibfnamefont{S.}~\bibnamefont{Lynch}},
  \bibinfo{author}{\bibfnamefont{M.}~\bibnamefont{Warner}},
  \bibinfo{author}{\bibfnamefont{A.}~\bibnamefont{Stoneham}},
  \bibinfo{author}{\bibfnamefont{G.}~\bibnamefont{Aeppli}},
  \bibinfo{author}{\bibfnamefont{D.}~\bibnamefont{Paul}}, \bibnamefont{et~al.},
  \bibinfo{journal}{Proc. Natl. Acad. Sci. USA} \textbf{\bibinfo{volume}{105}},
  \bibinfo{pages}{10649} (\bibinfo{year}{2008}).

\bibitem[{\citenamefont{Tyuterev et~al.}(2010)\citenamefont{Tyuterev, Sjakste,
  and Vast}}]{2P0CALC}
\bibinfo{author}{\bibfnamefont{V.}~\bibnamefont{Tyuterev}},
  \bibinfo{author}{\bibfnamefont{J.}~\bibnamefont{Sjakste}}, \bibnamefont{and}
  \bibinfo{author}{\bibfnamefont{N.}~\bibnamefont{Vast}},
  \bibinfo{journal}{Phys. Rev. B} \textbf{\bibinfo{volume}{81}},
  \bibinfo{pages}{245212} (\bibinfo{year}{2010}).

\bibitem[{\citenamefont{Pavlov et~al.}(2000)\citenamefont{Pavlov, Zhukavin,
  Orlova, Shastin, Kirsanov, H\"ubers, Auen, and Riemann}}]{THZLASER}
\bibinfo{author}{\bibfnamefont{S.~G.} \bibnamefont{Pavlov}},
  \bibinfo{author}{\bibfnamefont{R.~K.} \bibnamefont{Zhukavin}},
  \bibinfo{author}{\bibfnamefont{E.~E.} \bibnamefont{Orlova}},
  \bibinfo{author}{\bibfnamefont{V.~N.} \bibnamefont{Shastin}},
  \bibinfo{author}{\bibfnamefont{A.~V.} \bibnamefont{Kirsanov}},
  \bibinfo{author}{\bibfnamefont{H.-W.} \bibnamefont{H\"ubers}},
  \bibinfo{author}{\bibfnamefont{K.}~\bibnamefont{Auen}}, \bibnamefont{and}
  \bibinfo{author}{\bibfnamefont{H.}~\bibnamefont{Riemann}},
  \bibinfo{journal}{Phys. Rev. Lett.} \textbf{\bibinfo{volume}{84}},
  \bibinfo{pages}{5220} (\bibinfo{year}{2000}).

\bibitem[{\citenamefont{Morishita et~al.}(2009)\citenamefont{Morishita,
  Vlasenko, Tanaka, Semba, Sawano, Shiraki, Eto, and Itoh}}]{ITOHNMR}
\bibinfo{author}{\bibfnamefont{H.}~\bibnamefont{Morishita}},
  \bibinfo{author}{\bibfnamefont{L.~S.} \bibnamefont{Vlasenko}},
  \bibinfo{author}{\bibfnamefont{H.}~\bibnamefont{Tanaka}},
  \bibinfo{author}{\bibfnamefont{K.}~\bibnamefont{Semba}},
  \bibinfo{author}{\bibfnamefont{K.}~\bibnamefont{Sawano}},
  \bibinfo{author}{\bibfnamefont{Y.}~\bibnamefont{Shiraki}},
  \bibinfo{author}{\bibfnamefont{M.}~\bibnamefont{Eto}}, \bibnamefont{and}
  \bibinfo{author}{\bibfnamefont{K.~M.} \bibnamefont{Itoh}},
  \bibinfo{journal}{Phys. Rev. B} \textbf{\bibinfo{volume}{80}},
  \bibinfo{pages}{205206} (\bibinfo{year}{2009}).

\end{thebibliography}
\end{document}